\documentclass[12pt]{elsarticle}

\usepackage{enumitem}

\usepackage[english]{babel}
\usepackage[utf8x]{inputenc}
\usepackage{hyperref}
\usepackage{graphicx}
\usepackage{caption}
\usepackage[labelformat=simple]{subcaption}

\usepackage{mathtools}
\usepackage{xfrac}
\usepackage{siunitx}
\usepackage{placeins}
\usepackage{tikz}
\usepackage{pgfplots}
\usetikzlibrary{calc}
\usetikzlibrary{backgrounds}
\usepgflibrary{shapes}
\usetikzlibrary{through}
\usetikzlibrary{intersections}
\usetikzlibrary{matrix}
\usepackage{multirow}
\usepackage{hyperref}
\usepackage{cleveref}
\crefname{figure}{Fig.}{Figs.}
\Crefname{figure}{Figure}{Figures}

\crefname{table}{Table}{Tables}
\Crefname{table}{Table}{Tables}

\usepackage{xifthen}

\usepackage{esdiff}
\usepackage{ifthen}

\usepackage{booktabs}
\usepackage{nicefrac}

\usepackage[
]{todonotes}
\presetkeys{todonotes}{inline}{}
\usepackage{footmisc}

\usepackage[singlespacing]{setspace}

\usepackage{placeins}

\newcommand{\rz}{{\if mm {\rm I}\mkern -3mu{\rm R}\else \leavevmode
\hbox{I}\kern -.17em \hbox{R} \fi}}
\newcommand{\nz}{{\if mm {\rm I}\mkern -3mu{\rm N}\else \leavevmode
\hbox{I}\kern -.17em \hbox{N} \fi}}

\newcommand{\vv}[1]{\boldsymbol{#1}}

\newcommand{\x}{\ensuremath{\vv{\xi}}}

\newdimen\CdotAxis
\newcommand*{\CdotAux}[3]{%
  {%
    \settoheight\CdotAxis{$#2\vcenter{}$}%
    \sbox0{%
      \raisebox\CdotAxis{%
        \scalebox{#1}{%
          \raisebox{-\CdotAxis}{%
            $\mathsurround=0pt #2#3$%
          }%
        }%
      }%
    }%
    \dp0=0pt %
    \sbox2{$#2\bullet$}%
    \ifdim\ht2<\ht0 %
      \ht0=\ht2 %
    \fi
    \sbox2{$\mathsurround=0pt #2#3$}%
    \hbox to \wd2{\hss\usebox{0}\hss}%
  }%
}

\def\addlegendimage{\csname pgfplots@addlegendimage\endcsname}

\DeclareSIUnit{\molpc}{mol\text{-}\%}

\newcommand{\corM}[1]{\textcolor{black}{#1}}

\biboptions{sort&compress}

\journal{Computational Materials Science}
\begin{document}

\begin{frontmatter}
%
\author{Marco Seiz}
\ead{marco.seiz@kit.edu}

\address{Institute for Applied Materials (IAM), Karlsruhe Institute of Technology (KIT) Stra{\ss}e am Forum 7, 76131 Karlsruhe, Germany}

\title{Effect of rigid body motion in phase-field models of solid-state sintering}
%
\begin{abstract}
In the last two decades, many phase-field models for solid-state sintering have been published.
Two groups of models have emerged, with and without the contribution of rigid body motion.
This paper first describes the previously published phase-field model with an advection term driven by rigid body motion.
The model is then used to investigate the differences between models with and without rigid body motion in new benchmark geometries exhibiting markedly different behavior.
Sensitivity studies concerning the parameters of the rigid-body motion model are conducted and their effects on equilibrium and kinetic properties explored.
In particular, it is shown by simulations that a shrinkage rate independent of system size requires the inclusion of an advection term.
Finally, the reason behind this behavior is explored and implications for diffusion-only models are drawn.

\end{abstract}

 \begin{keyword}
 phase-field \hspace{1cm} solid-state sintering  \hspace{1cm} rigid body motion
 \end{keyword}

\end{frontmatter}

\section{Introduction}
The sintering process has been actively used by humans for millenia to manufacture products ranging from the humble coffee cup over spark plug insulators to complex printed electronics.
At the same time, it is also a naturally occurring process responsible for, among other things, snow compaction on mountains and glaciers\cite{Schleef2013}.
Thus a fundamental understanding of the process is essential in order to advance both manufacturing as well as predict the effect of climate change on glaciers.
The sintering process is based on the reduction of interface energies, either surfaces or grain boundaries, and is thus a spontaneous process occurring via diffusive mass transport.
Commonly six different pathways are identified: Volume diffusion from the surface to the grain boundary, volume diffusion from the bulk to the grain boundary, diffusion within the grain boundary, diffusion along the surface, vapor transport as well as plastic flow.
These transport pathways build the basis for geometric models of sintering such as the two-particle model\cite{Rahaman2003}.
While these are analytically solvable, they cannot be easily applied to real green bodies\cite{Exner1973} and thus more complex simulation methods need to be employed.

One such method is the phase-field method, which has been recently applied to the simulation of solid-state sintering in many different contexts.
Within each of these contexts, two approaches in modelling the solid-state sintering process have been established:
On the one hand there is a purely diffusive approach in which the phase-field model of choice is enhanced by incorporating diffusive pathways {\cite{Kumar2010,Hoetzer2019,Greenquist2020}}.
On the other hand there is an approach which, in addition to diffusive pathways, incorporates a kind of rigid body motion (RBM) by including an advective term in the evolution equations \cite{kazaryan1999generalized,Wang2006,Abdeljawad2019,Termuhlen2021,biswas2016study,biswasimplementation,Dzepina2019,Wang2021,Shi2021,Ivannikov2021}.
Some investigations \cite{Termuhlen2021,Shi2021} include a comparison of simulations with and without RBM, with the observed difference being a less pronounced shrinkage if no RBM is included.
However, this does not readily imply that RBM is a necessary ingredient for representing sintering, since a higher shrinkage at the same time could simply be achieved by an increase in the diffusion coefficients.
This paper proposes a new kind of benchmark geometry, a linear chain of grains, in which there is a marked difference between diffusion-only and coupled diffusion-RBM models which cannot be fixed by a simple change of parameters.
First, the classical RBM phase-field model of Wang \cite{Wang2006} will be introduced and a brief overview of the literature based on this model is presented.
Next, the benchmark geometry is detailed and simulations are conducted, with the parameter of interest being the shrinkage and how it is affected by the number of particles in the chain as well as the RBM model parameters.
Based on the simulation results conclusions in regards to the applicability of diffusion-only models and the correct parametrization of RBM model are drawn.


\section{Model}

For simplicity, the phase-field model described by Wang\cite{Wang2006} is employed, as it is often used as a base model for incorporating RBM.

The free energy functional $F$ is written as 
\begin{align}
F &= \int f(\rho, \vec{\eta}) + 0.5 \sum_\alpha  \beta_\eta |\nabla \eta_\alpha|^2   + 0.5 \beta_\rho |\nabla \rho|^2 dV \\
f(\rho, \vec{\eta}) &= A \rho^2 (1-\rho)^2 \nonumber\\
 &+ B \Big[ \rho^2 + 6(1-\rho)\sum_\alpha (\eta_\alpha^2) - 4 (2-\rho)\sum_\alpha (\eta_\alpha^3)  + 3 (\sum_\alpha \eta_\alpha^2)^2  \Big] 
\end{align}
with $f(\rho, \vec{\eta})$ accounting for the bulk energy of the individual phases and the remaining terms for the gradient energy associated with interfaces.
The phase-field vector $\vec{\eta} = (\eta_\alpha, \eta_\beta, \ldots, \eta_N)$ identifies individual grains and the density $\rho$ differentiates between the dense grains and the surrounding vacuum.
The bulk energy $f$ has minima corresponding to the $N$ grains as well as the surrounding vacuum.
A $\alpha$ grain phase is described by the state $\{\eta_\alpha = 1, \eta_{\beta \neq \alpha} = 0, \rho = 1\}$, and the vacuum phase by the state $\{\eta_\alpha = 0 \forall \alpha, \rho = 0\}$.
The mass density $\rho$ might be thought of as a relative measure of number density of atoms and thus $1-\rho$ could be considered a relative measure of vacancy density:
Low within the grains and high within the surrounding vacuum.
The bulk energy is a well type of potential and thus the phase-field profile has infinite extent, however with most of the gradients contained in a thin region.
The dynamics of the system follow based on the gradient flow of the nonconserved order parameter $\vec{\eta}$ and the conserved order parameter $\rho$ with an additional advection term:
\begin{align}
 \dot{\eta_\alpha} &= -L \frac{\delta F}{\delta \eta_\alpha} - \nabla \cdot (\eta_\alpha \vec{v_\alpha})\\
 \dot{\rho} &= \nabla \cdot ( D(\vec{\eta},\rho) \nabla \frac{\delta F}{\delta \rho} - \rho \vec{v})
\end{align}
The gradient flow construction of non-conservative $\eta_\alpha$ and conservative $\rho$ ensures that the coupled system of equations minimizes the free energy $F$ by following the variational derivative $\frac{\delta F}{\delta q} = \frac{\partial F}{\partial q} - \nabla \cdot \frac{\partial F}{\partial \nabla q}$ with $q$ describing the spatial function to be minimized.
Note that the advection term is not part of the free energy minimization and thus in effect plays the role of an outside force which could work against the free energy minimization.
The factor $L$ is a mobility controlling how quickly the non-conserved order parameters $\eta_\alpha$ relax to equilibrium.
The mass diffusivity $D$ is formulated as
\begin{align}
   D(\vec{\eta},\rho) &= D_{vol}\phi(\rho) + D_{vap}(1-\phi(\rho)) \\ 
  &+D_{surf} \rho (1-\rho) + D_{gb} \sum_\alpha \sum_{\beta \neq \alpha} \eta_\alpha \eta_\beta \nonumber\\
  \phi(\rho) &= \rho^3 (10 - 15\rho + 6\rho^2)
\end{align}
following Wang in his original description. 
This incorporates the effects of diffusion within both the solid and the surrounding vacuum as well as the enhanced diffusion on the surface and grain boundaries.
While Wang gave no motivation for the choice of interpolation functions, let us consider these briefly:
The volume diffusion is interpolated with $\phi(\rho)$ which has the properties $\phi(0) = 0$ and $\phi(1) = 1$.
This ensures that vapor diffusion only occurs in the vapor ($\rho = 0$) and volume diffusion only within the grains ($\rho = 1$).
The same applies for the formulations for surface and grain boundary diffusion, as these vanish on reaching their respective bulk regions as well.

Following Wang, an effective force density acting on the grain boundary of a grain $\alpha$
\begin{align}
  \vec{dF_\alpha} &= \kappa \sum_{\beta \neq \alpha} (\rho-\rho_{gb})g(\alpha,\beta) (\nabla {\eta_\alpha} - \nabla {\eta_\beta})
\end{align}
is postulated.
This formulation is motivated by the fact that grain boundaries exhibit a lower mass density ($\rho_{gb}$) than the accompanying bulk material ($\rho_{eq}=1$) and thus should exert a force on neighboring grains in order to achieve this density.
The difference of phase-field gradients ensures conservation of momentum for a constant $\rho_{gb}$ since for a grain boundary $\alpha\beta$, the grain $\alpha$ will be affected by an equal but opposite force density as its neighboring grain $\beta$.
The parameter $\kappa$ is a stiffness relating the force magnitude to the density deviation within the grain boundary.
The filtering function $g$ constrains the force density to the grain boundary itself:
\begin{align}
   g(\alpha,\beta) &= \begin{cases}
                     1, & \eta_\alpha \eta_\beta \geq c\\
                     0, & else\\
                    \end{cases} 
\end{align}
since only in grain boundaries the product $\eta_\alpha \eta_\beta$ can be larger than a $c > 0$.
The force density is integrated for each grain in order to determine a resultant force per grain:
\begin{align}
 \vec{F_\alpha} &= \int_V \vec{dF_\alpha} dV
\end{align}
If the force density is asymmetric with respect to the grain boundary center, there would also be a resultant torque.
This torque is not considered in this paper for two reasons:
First, the geometries considered within are always symmetric with respect to the grain boundary center and hence any resultant torque can only originate from numerical errors.
Second, this resultant torque does not contribute to densification following Shi et al. \cite{Shi2021}; since this paper is concerned with densification, dropping the torque is reasonable as well.

The resultant force is now assumed to cause an instantaneous movement of the entire grain as a rigid body with its velocity described by
\begin{align}
   \vec{v_{\eta_\alpha}} &= \frac{m_t}{V_\alpha} \vec{F_\alpha}\\
  V_\alpha &= \int_V \eta_\alpha dV
\end{align}
and hence the grains are moved towards or away from their grain boundary, depending on the orientation of $\vec{F_\alpha}$.
The parameter $m_t$ is a translational mobility and the volume $V_\alpha$ incorporates the particle size, with larger particles being moved more slowly than smaller ones at equivalent force.

This velocity is locally interpolated with the phase-field $\eta_\alpha$ in order to determine the local velocity for the individual phase-fields, with the sum of velocities describing the velocity of the mass density field $\rho$:
\begin{align}
  \vec{v_\alpha}(\vec{x}) &=  \vec{v_{\eta_\alpha}} \eta_\alpha(\vec{x})  \label{eq:valpha}\\
  \vec{v}(\vec{x}) &= \sum_\alpha \vec{v_\alpha}(\vec{x})  \label{eq:velocity}
\end{align}
The interpolation with $\eta_\alpha$ ensures a smooth transition between grains of different velocity.

The parameters appearing in all equations are the same as in Wang's original paper \cite{Wang2006} unless mentioned otherwise.
Following the analysis by Ahmed et al.\cite{Ahmed2013}, the free energy parameter set $\{A = 16, B = 1, \beta_\eta = 1, \beta_\rho = 10\}$ corresponds to values of the surface energy $\gamma_s = \frac{23 \sqrt{3}}{18}$, the grain boundary energy $\gamma_{gb} = \frac{2 \sqrt{3}}{3}$ and the interface width $W = \frac{2 \sqrt{3}}{3}$.
The equilibrium dihedral angle given by $\psi = 2\arccos(\frac{\gamma_{gb}}{2\gamma_s})$ is about \SI{150}{\degree}.
The diffusion coefficients are assumed to be $D_{surf}=4, D_{gb}=0.4, D_{vol}=0.01$ and $D_{vap}=0.001$ and the mobility $L=10$.
The parameters $m_t = 500$ and $\kappa = 100$ control the relationship between the density deviation within the grain boundary and the resulting velocity of each grain.
The evolution equations for the phase-fields and the density are evaluated using a standard forward-time-central-space (FTCS) scheme. 
Since no details on the discretization parameters were provided by Wang, a spatial discretization of $\Delta x = 0.33$ was chosen, which led to a stable timestep size of $\Delta t = 2.9648025\cdot10^{-5}$.
As in the original paper, the parameters do not correspond to a specific material system or thermodynamic conditions.
The results are thus generic for material systems exhibiting similar surface energy and diffusion ratios as the ones employed here.

A brief review of the similarities and differences in papers concerning rigid-body motion in sintering is given in the following:
In the first paper on the solid-state sintering problem of Biswas et al. \cite{biswas2016study} the rigid-body motion terms are adapted verbatim from Wang's original paper, as is done in the present work.
In the second paper \cite{biswasimplementation} the approach slightly changed, specifically the local velocity $v_\alpha$ from \cref{eq:valpha} is no longer interpolated with the order parameter $\eta_\alpha$.
Abdeljawad et al. \cite{Abdeljawad2019} simplified the model by combining the parameters $\kappa$ and $m_t$ as these only appear as products in the final evolution equation.
Additionally the rotational velocity was assumed to vanish.
Dzepina et al. \cite{Dzepina2019} were the first to deviate significantly from Wang's original formulation:
A pairwise force $F_{ij} = \vec{N} (c(t_0) - c(t))$ is defined, with $\vec{N}$ being the connecting vector between the center of mass of a particle $i$ and the center of the contact point, and $c(t)$ being the density at the center of the contact area between grains $i$ and $j$ at the time $t$.
It is assumed that at time $t_0$ the density or equivalently vacancy concentration is in equilibrium, such that the force acts as a spring keeping the particles connected.
The force is then connected to the velocity via $v_{i} = \frac{m_t}{V_i} \sum_{j!=i} F_{ij}$.
Furthermore, the influence of an external pressure is included in the evolution of the order parameters.
Termuhlen et al. \cite{Termuhlen2021} adapted the formulation of Wang verbatim, but improved the implementation such that an order parameter reassigning scheme works correctly with the RBM terms.
This allowed them to simulate up to 3000 particles with only 34 different order parameters.
In a smaller simulation with 332 particles, they also conducted a set of simulation with and without rigid body motion and could show that the incorporation of rigid-body motion increased the densification rate.
Wang et al. \cite{Wang2021} considered the laser sintering process in which not only solid-state sintering takes place but also local melting and re-solidification.
The solid-liquid phase transformation is included by a temperature dependence of the free energy density.
The advection velocity for the density field includes both the rigid-body motion as well as the melt flow.
The RBM terms are again taken verbatim from Wang's original paper.
Shi et al. \cite{Shi2021} employed the RBM model verbatim from Wang in a three-dimensional three particle setup, creating a pore between the three particles whose evolution is related to densification.
In contrast to earlier works, the free energy in terms of the phase-field was modified to be rather similar to the model proposed by Steinbach et al. \cite{Steinbach1999}, with a modified gradient energy and an obstacle type potential.
Furthermore, the evolution equation of the phase-field is based on the idea of interface fields, also suggested by Steinbach et al. \cite{Steinbach1999}.
Similar to the present study, Shi et al. investigated the effect RBM has on densification.
They could show that without RBM densification happens at a slower rate.
Furthermore, they showed that the torque has no effect on densification and that the parameter $\kappa$, once above a certain threshold, has little effect on densification rate.
However, they did not investigate the influence of the parameter $\rho_{gb}$.
Ivannikov et al. \cite{Ivannikov2021} deviated significantly from Wang's original formulation.
Their formulation is based on the free energy change due to particle motion, which should achieve a minimum in equilibrium.
Enforcing this constraint leads to a displacement $\Delta s$ towards the grain boundary $\Delta s = 2 \int (\rho_{gb} w - c)w dA_{gb} / (\int \nabla \eta_2 - \nabla \eta_1) w dAgb$ for a two-particle system, with a weighting function $w=\eta_1 \eta_2$ and the differential area of the grain boundary $dA_{gb}$.
This is translated into an advection velocity via $v = \frac{\Delta s}{\Delta t}$ with the timestep $\Delta t$.
Note that this formulation does not require specification of the stiffness $\kappa$ and mobility $m_t$ which were still free parameters within Wang's original formulation.

For the behavior of the model in comparison to the classical two-particle system and its accompanying power laws, the reader is referred to the existing literature \cite{Wang2006,biswas2016study,biswasimplementation}.

\section{Benchmark geometry and analysis methods}

\begin{figure}[htb]
\centering
 \includegraphics[width=\columnwidth]{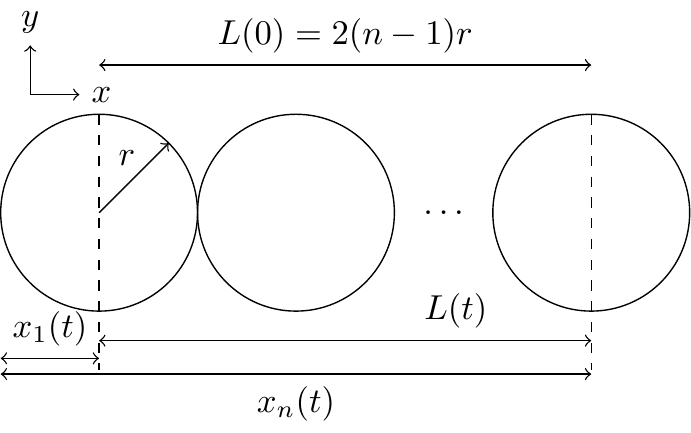}
 \caption{Finite, linear chain of $n$ particles of radius $r$.
 The coordinate $x_i$ describes the barycenter of the $i$th particle along the chain axis.
 The length of the chain is represented by the center-to-center distance of the first and last particle.}
 \label{fig:chain}
\end{figure}

A finite, linear chain of equally-sized grains is generally considered in this paper.
For circular grains \cref{fig:chain} shows a geometric sketch along with the relevant parameters of the number of grains $n$ and the radius $r$.
During sintering vacancies are annihilated at the grain boundary, which yields both neck growth and densification.
Assuming that each grain boundary acts independently and absorbs the same amount of vacancies, one would expect that the densification rate at a certain time is independent of the number of grain boundaries in the chain.
Based on this reasoning, the densification is taken to be the main parameter of interest in this study, as it should stay invariant with the number of particles $n$, i.e. the densification-time curves should form a single master curve regardless of $n$.
As a measure of densification the strain $\epsilon(t)$ is used, computed by comparing the distance of barycenters of the leftmost ($x_1$) and rightmost ($x_n$) particle (cf. \cref{fig:chain})
\begin{align}
 L(t)     &= x_{n}(t) - x_{1}(t) \\
 \epsilon(t) &= \frac{L(t) - L(0)}{L(0)} \\
 &= \frac{\Delta L}{L(0)} 
\end{align}
with the x-coordinate being the linear direction of the particle chain.
The strain is positive if the chain lengthens and negative if the chain shrinks.
For the simulations with RBM, the individual displacements $u_i = v_i\Delta t$ only due to RBM were also tracked and integrated over time.
The length change calculated purely by these advection steps did not differ much from the length change calculated based on the barycenter movement.
Their time evolution was highly similar, with the barycenter distance method showing a slightly larger length change due to including the effects of diffusive transport.

The case of an infinite linear chain has previously been investigated via geometric models by several authors \cite{Kellett1989,Parhami1999,Cannon1989a}.
If one simply applies periodic boundary conditions to a two-particle geometry \corM{with the grains being cut by the periodic boundary}, then the net velocity as calculated by the model above will always vanish.
Thus a direct comparison with these is not possible.
However, a common point in these analyses is whether the geometry can be considered densifying or not, which leads to different equilibrium shapes.
Specifically, Kellet and Lange\cite{Kellett1989} showed for an infinite chain of cylinders that their equilibrium shape could be described with three variables, viz. the equilibrium dihedral angle $\psi$, the grain boundary length $h$ and the radius $r$ of the truncated sphere connecting two grain boundaries.
In the densifying case, the non-dimensional equilibrium radius $R=r/r_i$ is given by $R_{eq}^d = (\frac{\pi}{\pi - \psi + \sin(\psi)})^{1/2}$ whereas in the non-densifying case it is given by $R_{eq}^{nd} = \frac{1}{\cos(\psi / 2)}$, with the initial radius $r_i$.
In both cases, the non-dimensional grain boundary length $H = h/r_i$, or twice the neck radius $X$, is given by \corM{$H = \frac{\frac{\pi}{R} + R{(\psi - \pi + \sin(\psi)) }}{ (2 \cos(\psi/2))}$}.
Based on the resulting geometry, the shrinkage strain in equilibrium can be computed via $\epsilon = 1- R \cos(\psi/2)$, which yields 0.5466 for $\psi = \ang{150}$.

\begin{figure}[htb]
 \centering
 \includegraphics[width=\columnwidth]{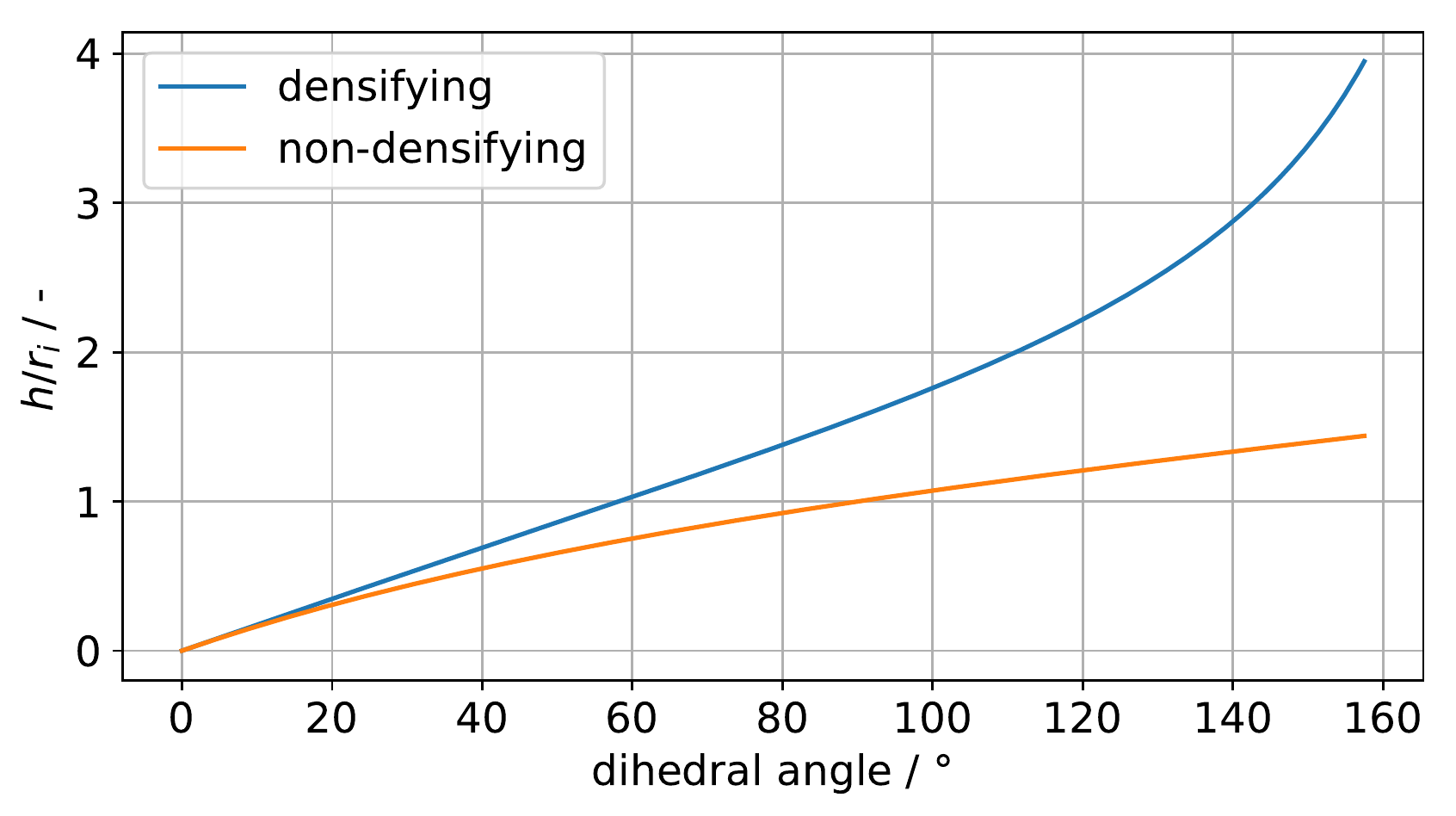}
 \caption{
 Equilibrium grain boundary length $h$ normalized by the initial grain radius $r_i$ over the equilibrium dihedral angle $\psi$.
 For increasing dihedral angle the grain boundary gets longer, with the densifying geometry generally exhibiting longer grain boundaries than the non-densifying geometry.
 }
 \label{fig:gblength}
\end{figure}

Plotting the grain boundary length over the dihedral angle yields \cref{fig:gblength}, in which it is easy to see that densifying geometries generally yield longer grain boundaries than non-densifying ones.
Specifically for a dihedral angle of $\ang{150}$ the non-densifying grain boundary length should be about $1.39r_i$ and the densifying one $3.38r_i$.
It seems reasonable that this result transfers to finite chains, with some error induced by the end particles taking on a different shape.
Thus if a system is densifying, one would expect much longer grain boundaries than in a non-densifying system, which allows the classification of phase-field models according to whether they describe a densifying geometry or not.
In order to calculate the grain boundary length $h$ the grain boundary area $A$ is divided by the interface width $W$.
\corM{The neck length then follows as $X=h/2$.}
The grain boundary area between the grains described by order parameters $\eta_\alpha$ and $\eta_\beta$ can be computed based on the phase field by $A = \int 4\eta_\alpha \eta_\beta dV$.

%


\section{Results}
\subsection{Chain length and densification}
\begin{figure}[hbt!]
\begin{center}
\begin{subfigure}[]{0.45\columnwidth}
  \includegraphics[width=\textwidth]{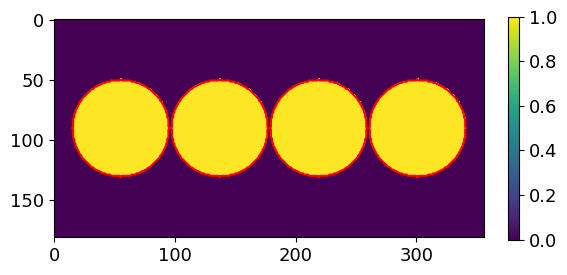}
  \caption{initial sharp interface configuration}
  \label{fig:sharp}
\end{subfigure}
\begin{subfigure}[]{0.45\columnwidth}
  \includegraphics[width=\textwidth]{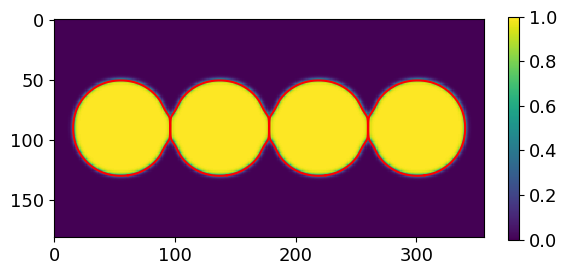}
  \caption{shortly before RBM activation}
  \label{fig:rbm}
\end{subfigure}
\begin{subfigure}[]{0.45\columnwidth}
  \includegraphics[width=\textwidth]{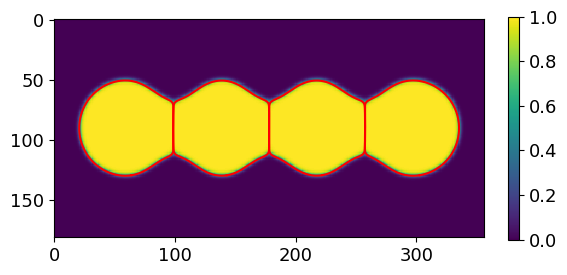}
  \caption{state at normal end of simulation,\\  $t=4\cdot10^5 \Delta t $}
  \label{fig:normal_end}
\end{subfigure}
\begin{subfigure}[]{0.45\columnwidth}
  \includegraphics[width=\textwidth]{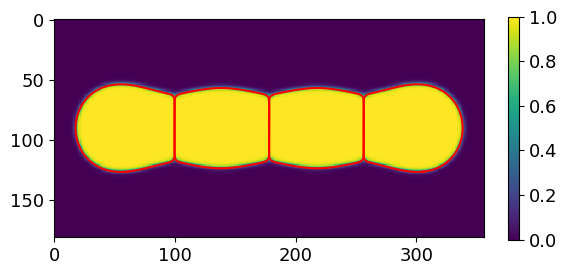}
  \caption{state at long end of simulation,\\  $t=3.6\cdot10^6 \Delta t $}
  \label{fig:long_end}
\end{subfigure}
  \caption{Time evolution of the sum of grain phases $\sum \eta_\alpha$ in a 4 particle chain with rigid body motion activated after $t=10^4 \Delta t$.
  The $\eta_\alpha=0.5$ contour lines of individual grains are drawn as red lines.
  From the start (a) to the activation of RBM (b) no significant densification is observed, but a neck is formed.
  At the regular end of the simulation (c), a significant densification is observed relative to the initial configuration (a).
  If this simulation is continued then material apparently flows towards the boundary (d) instead of the center of the chain.
}
  \label{fig:4pRBM}
  \end{center}
\end{figure}

The initial radius for the circular particles is chosen to be $r =$ 40 cells, with at least $20$ cells left empty between the particles and the grain boundary.
Chain lengths of 2, 4, 6 and 8 particles are considered, with and without RBM.
For this study, the RBM parameter set $\{\rho_{gb}, \kappa\}$ is kept fixed at $\{\rho_{gb} = 0.9816, \kappa=100\}$ as in the original paper by Wang.
On the boundary gradient-zero conditions for all phase-fields $\eta_\alpha$ and the density $\rho$ are employed.
In order to avoid pairing of particles due to the natural boundary effect of the first and last particle, the first $10^4$ time steps of $4\cdot10^5$ are calculated without RBM.
After the first $4\cdot10^5$ steps, corresponding to a simulation time of 11.8592, the simulations were analyzed and continued for another $32\cdot10^5$ steps in order to start investigating the long-time behavior.
The time evolution of a 4 particle chain which is continued with RBM is shown in \cref{fig:4pRBM}.
Both neck growth and shrinkage are observed, with an apparent unshrinkage being observed from (c) to (d) in the long-term simulation.

\begin{figure}[htb!]
 \centering
    \begin{subfigure}[]{0.48\columnwidth}
    \centering
        \includegraphics[width=\textwidth]{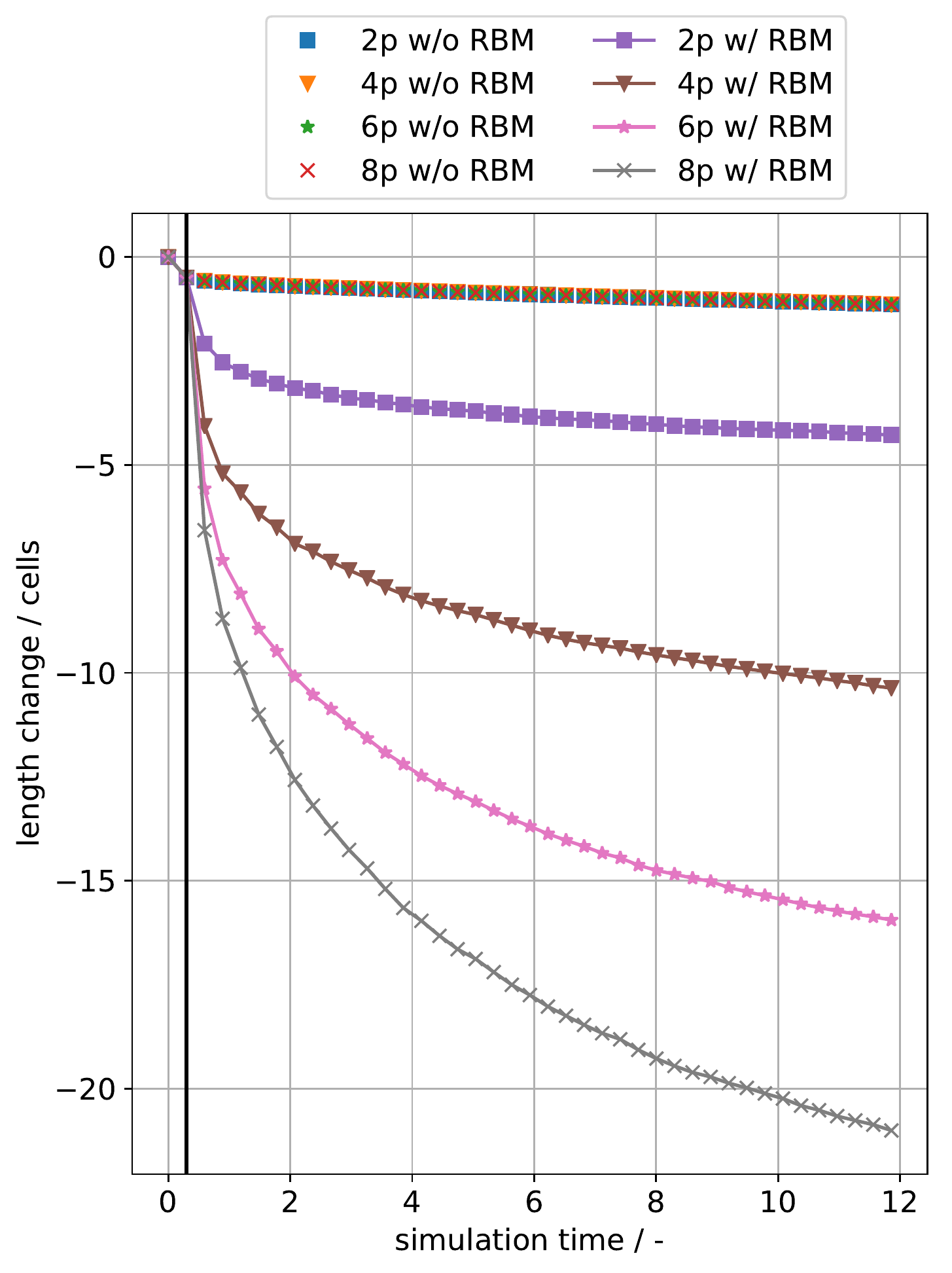}
        \caption{length change}
        \label{fig:lengthchange}
    \end{subfigure}
    \begin{subfigure}[]{0.48\columnwidth}
    \centering
        \includegraphics[width=\textwidth]{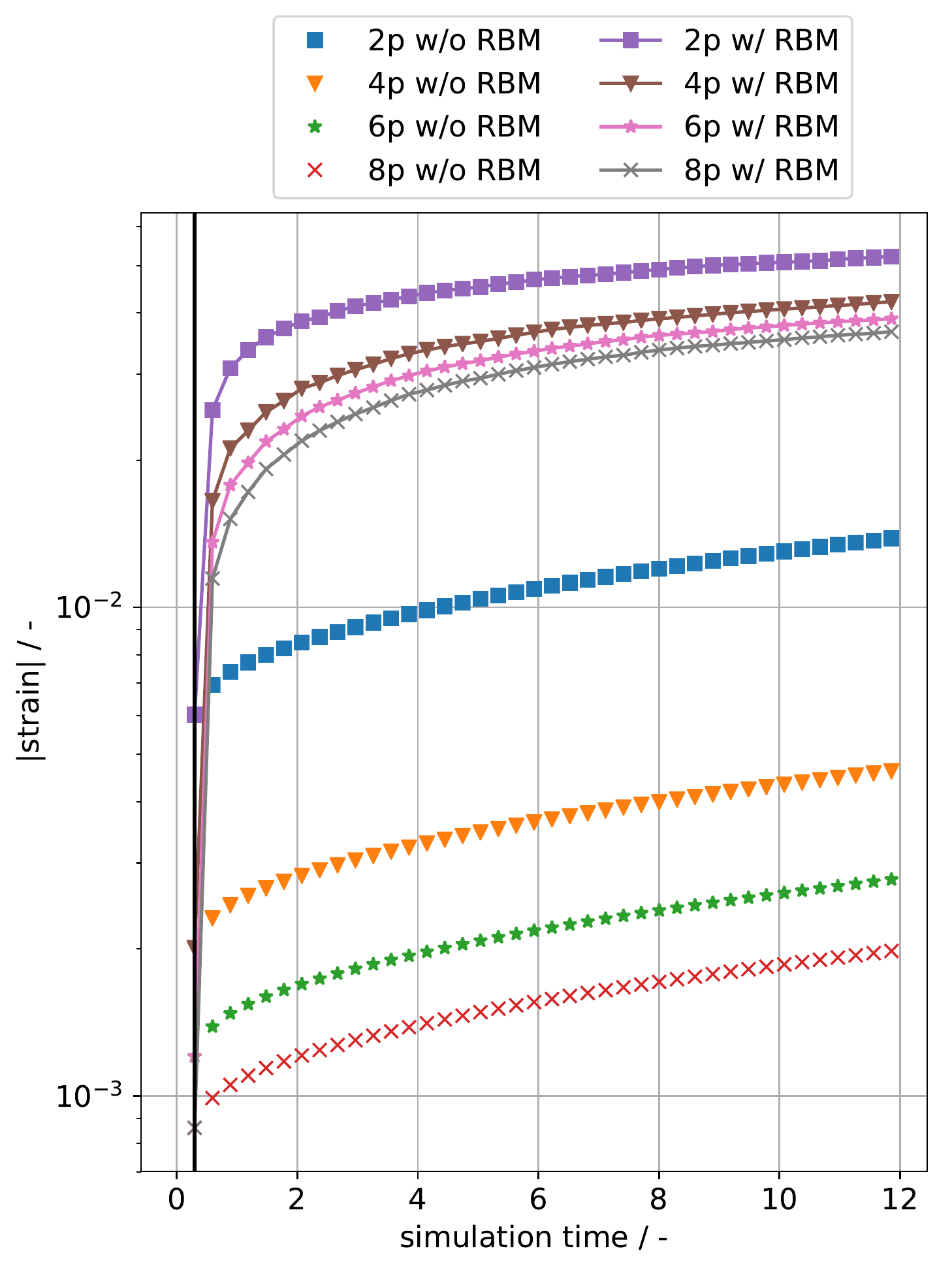}
        \caption{strain}
        \label{fig:densification}
    \end{subfigure}
 \caption{
Length change and absolute value of the strain for particles chains of various lengths with and without RBM.
Lines with markers indicate simulations with RBM, whereas only markers indicate simulations without RBM.
The label describes how many particles were in the chain as well as whether RBM was active.
The black vertical line indicates the time at which RBM was activated.
The strain is plotted on a semilogarithmic scale for better visibility of the differences without RBM.
Simulations without RBM show a length change independent of chain length, whereas simulations with RBM have an almost linear increase in length change with increasing chain length.
Thus the strain is variable for simulations without RBM and almost constant for those with RBM.
}
 \label{fig:densi}
\end{figure}

\FloatBarrier
The length change $\Delta L$ of the sample and the absolute value of the strain $|\epsilon(t)|$ are plotted in \cref{fig:densi}.
For the purely diffusive simulations the length change $\Delta L$ at any particular time is almost independent of chain length and thus a variable strain is observed.
Incorporating RBM yields increased length changes with increasing chain size, leading to almost the same strain being observed at any particular time for more than two particles.
This suggests that the the strain rate, or equivalently densification rate, does \emph{not} depend on system size if RBM is included.
Note that even with the inclusion of RBM there is a slight decrease of strain with chain length.



\begin{figure}[htb!]
  \centering
    \begin{subfigure}[]{0.4\columnwidth}
    \centering
        \includegraphics[width=\textwidth]{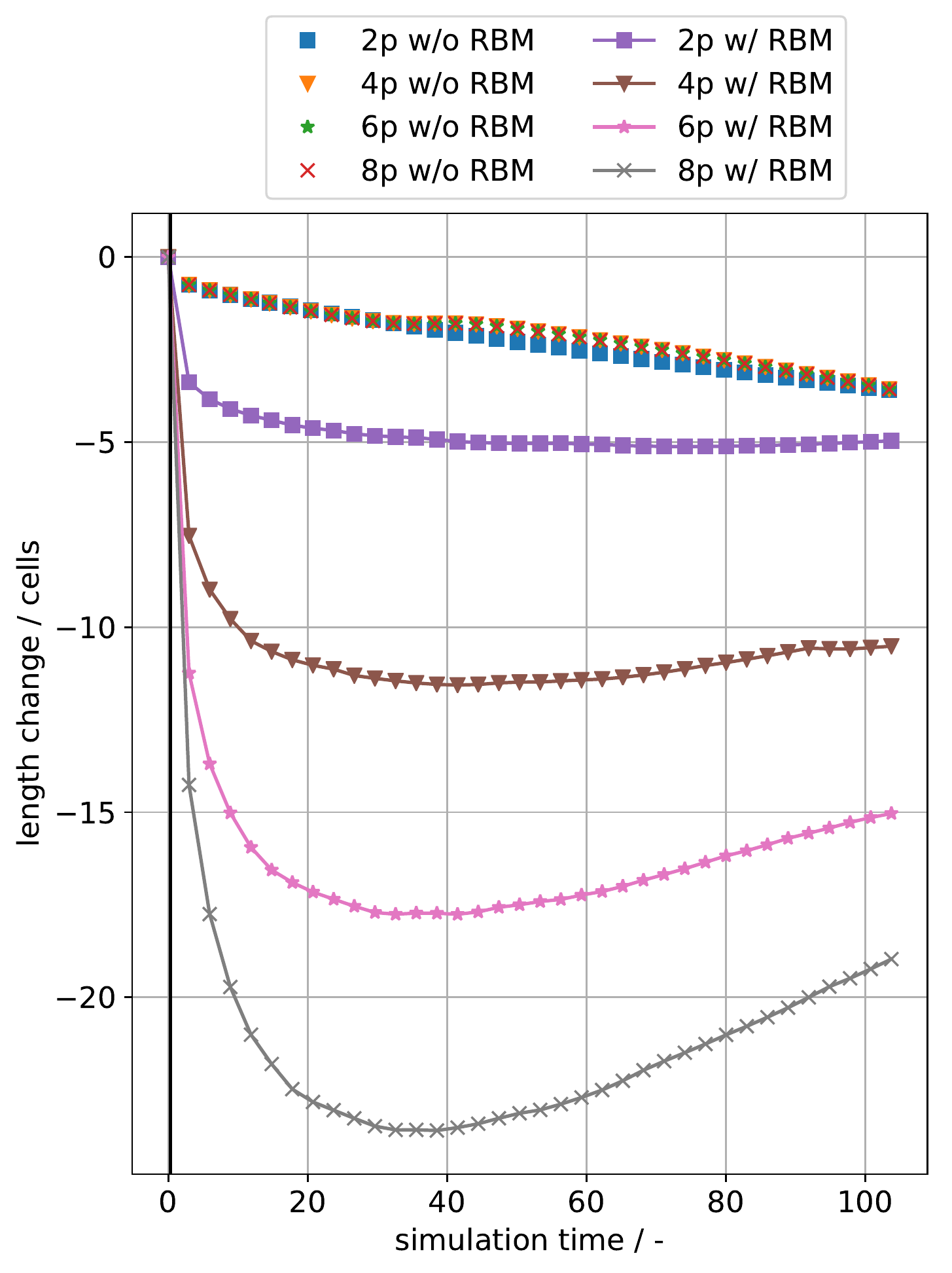}
        \caption{length change}
        \label{fig:lengthchange-long}
    \end{subfigure}
    \begin{subfigure}[]{0.4\columnwidth}
    \centering
        \includegraphics[width=\textwidth]{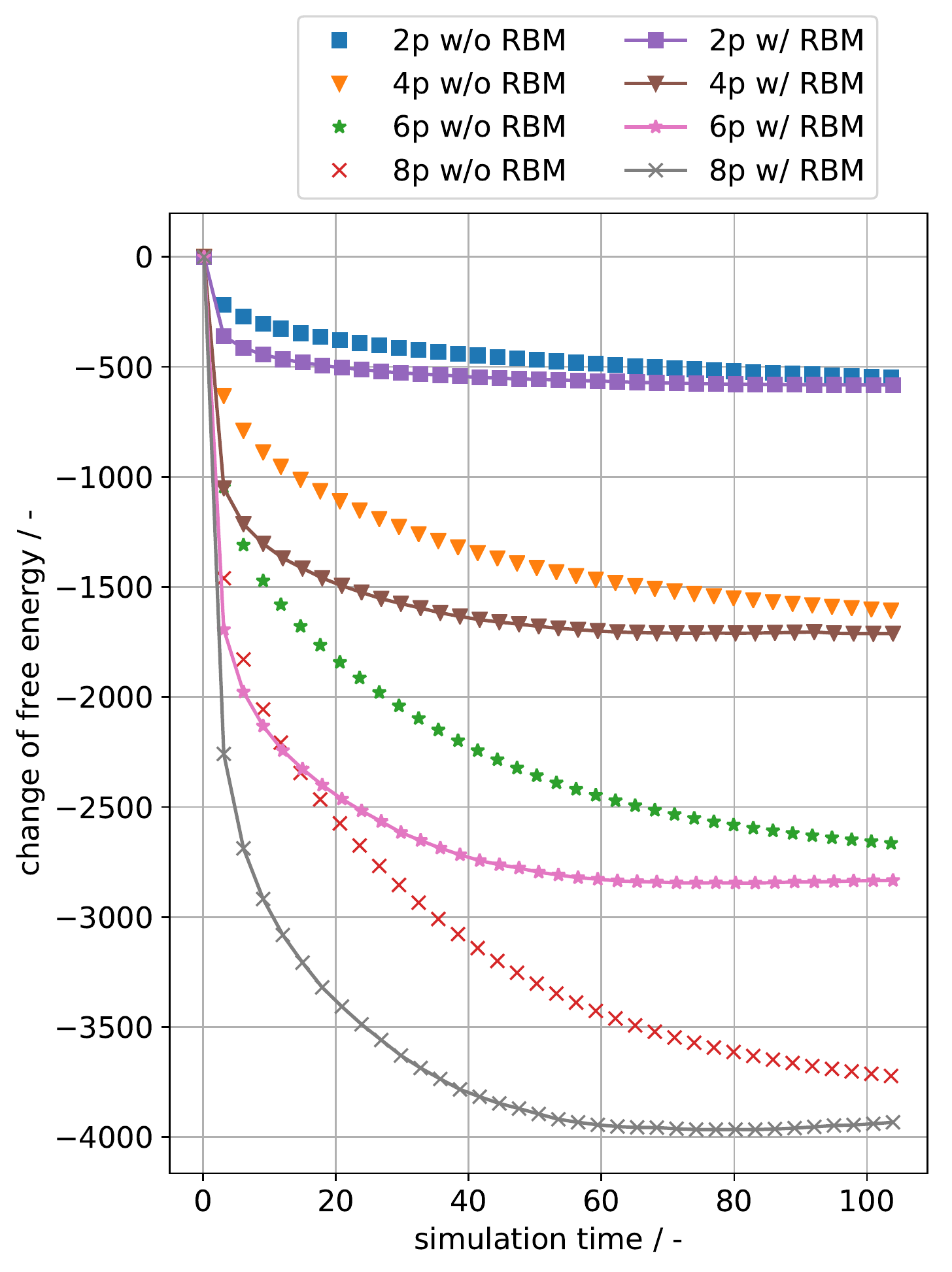}
        \caption{change of free energy after $t=2000\Delta t$}
        \label{fig:energy}
    \end{subfigure}
 \caption{Long time behavior of simulations with and without RBM.
 Simulations with RBM eventually show unphysical unshrinkage.
 This also causes the free energy to increase which is inconsistent with the minimization of free energy.
 }
 \label{fig:badRBM}
\end{figure}

\begin{figure}[htb!]
  \centering
  \includegraphics[width=0.8\columnwidth]{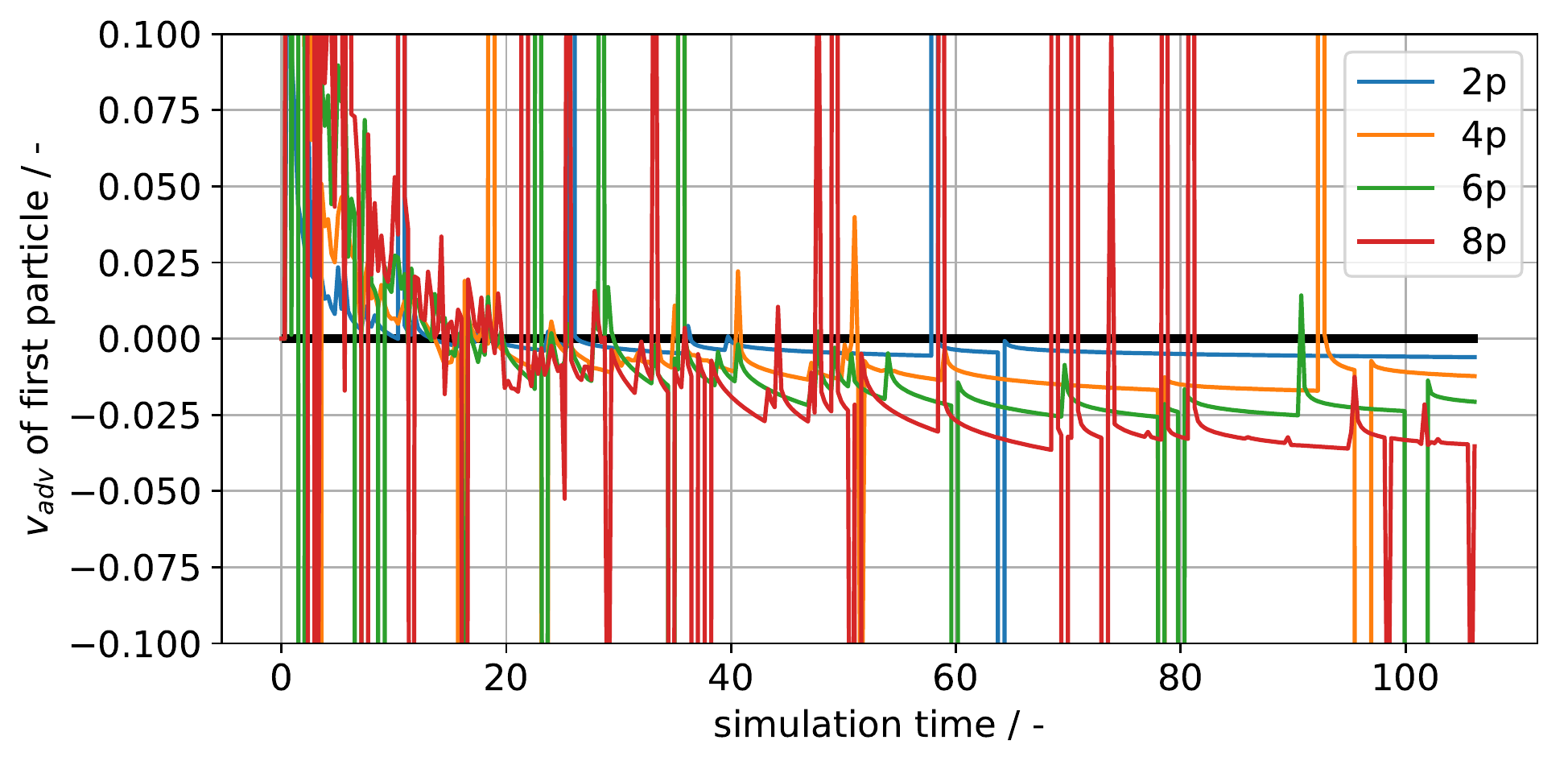}
  \caption{  
  \corM{The rigid-body velocity of the leftmost particle over time.}
  A positive velocity point towards the total center of mass and thus causes densification.
  While it is initially positive and thus densifying, the velocity becomes negative for extended periods of time during later stages, thus causing the observed unshrinkage.
  \corM{Since the velocity is calculated instantaneously based on the resulting force it has discontinuous jumps due to the filtering function $g$.}
  }
  \label{fig:velocity}
\end{figure}

These results suggest that RBM is indeed a necessary ingredient for a physically sensible phase-field model of sintering, as the kinetic pathway taken should obviously not depend on the system size.
However, the manner in which the rigid body velocity of each particle is calculated is also of great import.
This is revealed by looking at the long-term simulations, in which eventually a kind of unshrinkage occurs, shown in \cref{fig:badRBM}, which is accompanied by an increase in free energy.
This is due to the phase-specific velocities $v_{\eta_\alpha}$ no longer being oriented towards the total center of mass, but rather away from it.
Once the velocity points outwards it will transport mass towards the boundary and unshrinkage can occur.
The instantaneous velocity of the leftmost particle for the 4 simulations with RBM is shown in \cref{fig:velocity}, with a positive velocity pointing towards the total center of mass and a negative one away from it.
The view is restricted to the dimensionless velocity range of $[-0.1, 0.1]$ in order to emphasize the occurrence of long periods with negative velocities; these are the cause behind the observed unshrinkage.
\corM{The jumps of the instantaneous velocity are caused by the filtering function $g$, since new cells with large force densities are added to the resultant force in an abrupt manner.
Note that this particle velocity is used for the spatial interpolation (\cref{eq:valpha}) and hence does not need to be continuous.
}

\begin{figure}[htb!]
 \centering
    \begin{subfigure}[]{0.48\columnwidth}
    \centering
        \includegraphics[width=\textwidth]{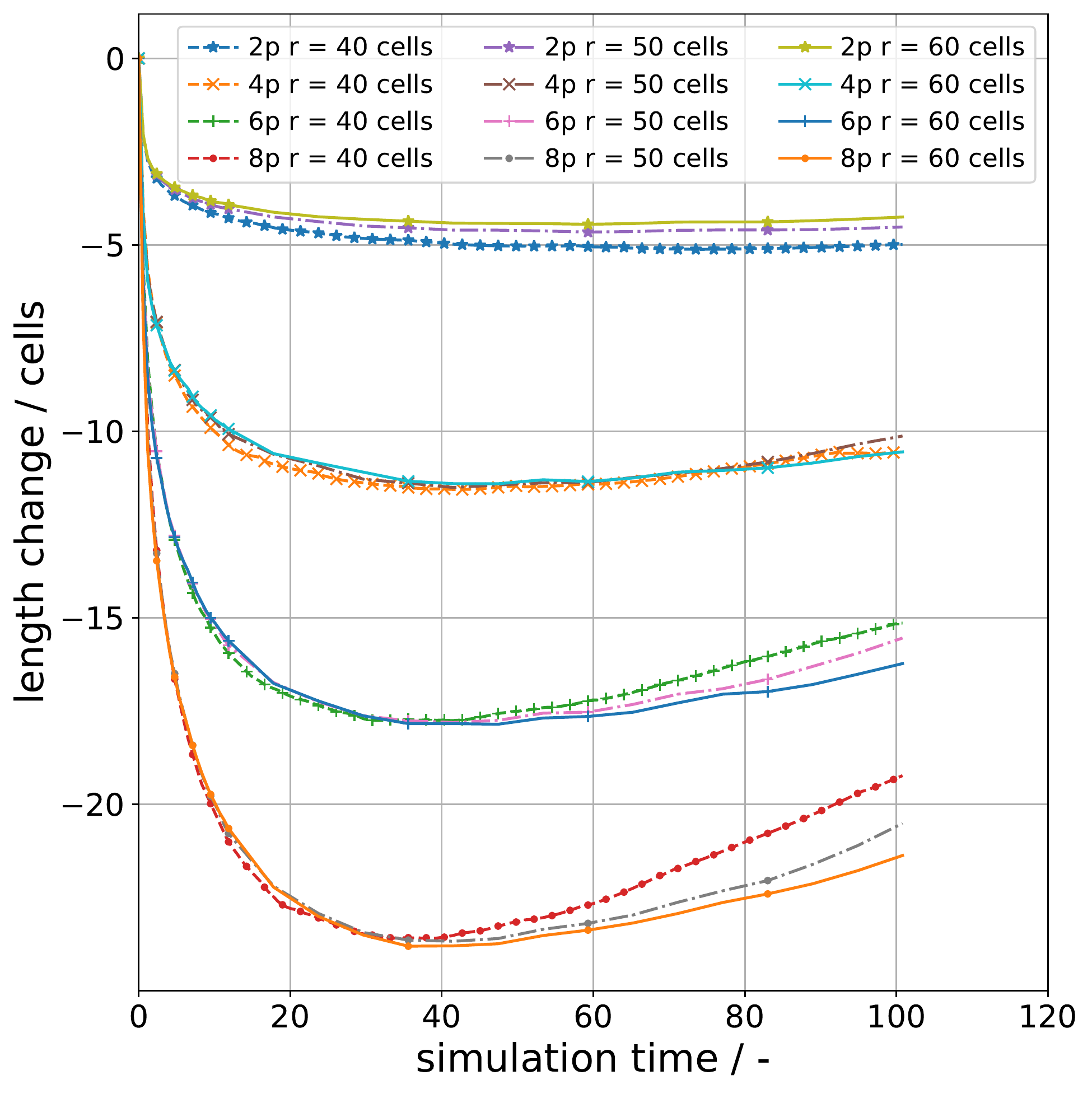}
        \caption{length change}
        \label{fig:dl-rvar}
    \end{subfigure}
    \begin{subfigure}[]{0.48\columnwidth}
    \centering
        \includegraphics[width=\textwidth]{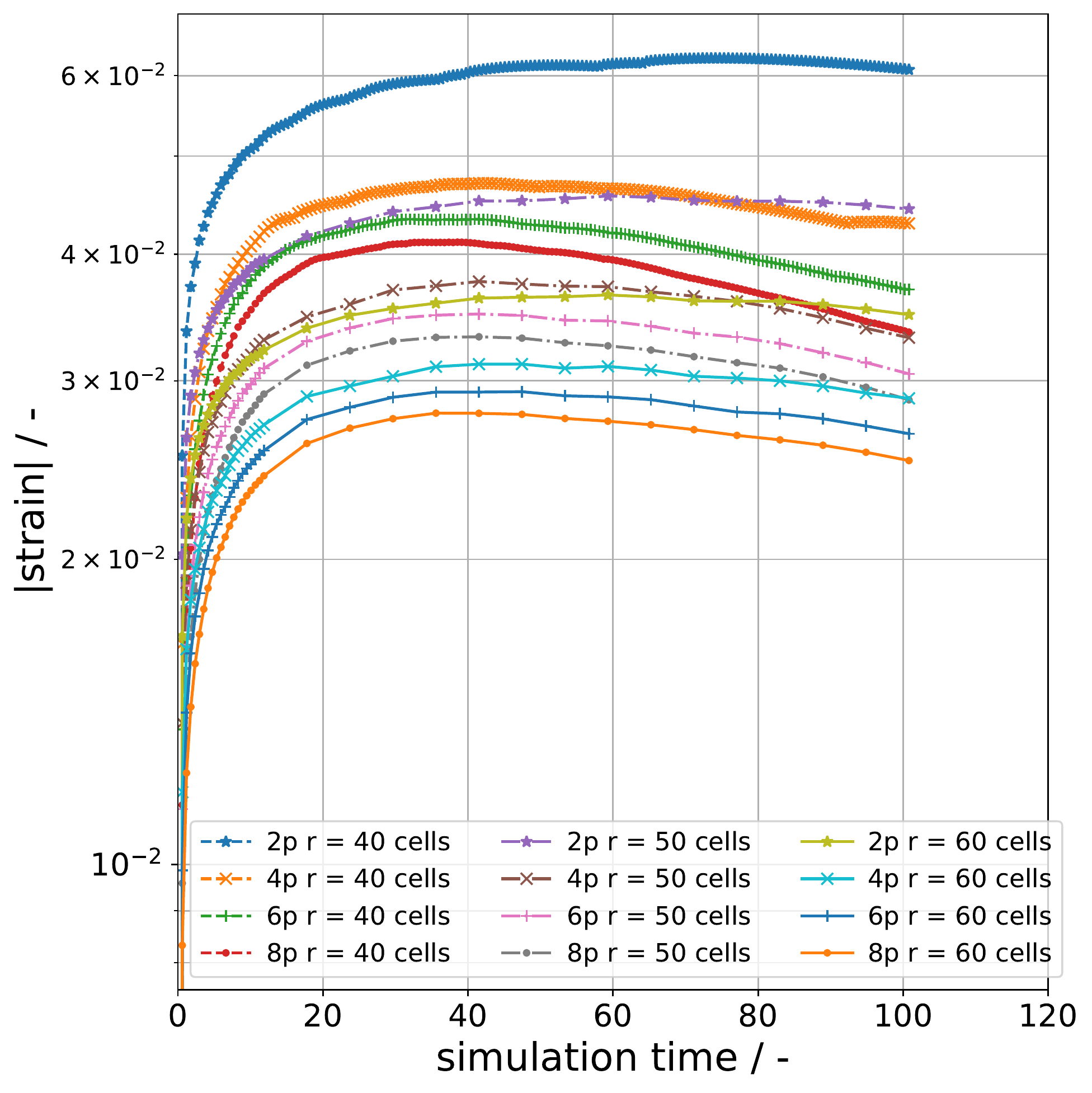}
        \caption{strain}
        \label{fig:shrink-rvar}
    \end{subfigure}
 \caption{
\corM{Length change and absolute value of the strain for particles chains of various lengths and particle sizes with RBM.
The length change is barely affected by the particle size, but this induces a large variation in observed strain.}
}
 \label{fig:rvar}
\end{figure}

\begin{figure}[htb!]
  \centering
  \includegraphics[width=0.8\columnwidth]{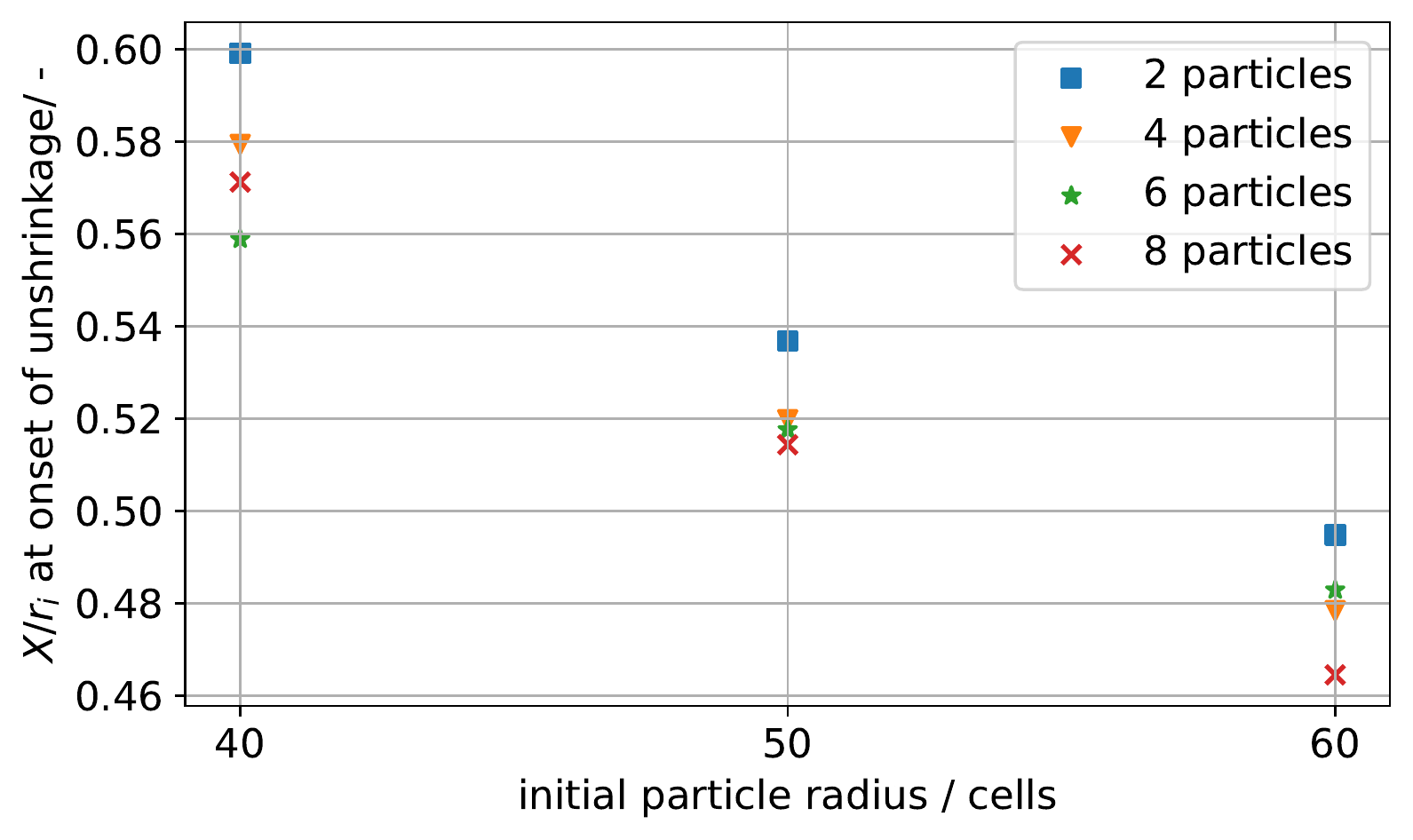}
  \caption{  
  \corM{The geometrical state, described by the relative neck radius $X/r_i$, at the onset of unshrinkage for the conducted simulations.
  Unshrinkage is observed at relatively smaller necks as particle size is increased, or when the number of particles in the chain is increased.
  However, the latter influence is not observed to be monotonic.}
  }
  \label{fig:geom_unshrink}
\end{figure}

\corM{Before proceeding to investigate the reason for the sign change in the velocity, the influence of particle size is considered shortly.
Two additional particle sizes, 50 and 60 cells, were simulated for chains of length 2, 4, 6 and 8 thus yielding another 8 simulations.
Their length changes and strains are depicted in \cref{fig:rvar}.
The length change is barely affected by the change in particle size, which in turn causes the densification to decrease as the particle size is increased.
The larger particles tend to experience less unshrinkage in total.}
\corM{
In order to determine whether particle size influences when unshrinkage starts the relative neck radius $X/r_i$, i.e. the neck radius divided by the initial particle radius, is evaluated at the onset of unshrinkage.
The onset of unshrinkage is assumed to be the global minimum of the length change curve.
\cref{fig:geom_unshrink} shows the dependence on the particle size as well as the number of particles in the chain.
The relative neck radius at the start of unshrinkage tends to decrease as the particle size is increased.
Furthermore, more particles in the chain seem to also incentivize unshrinkage, but not in a strictly monotonic way as the particle size.
}

In the following section the reason for the sign change in the velocity will be determined with theoretical considerations as well as sensitivity studies on the RBM parameters.

\FloatBarrier

\subsection{Shrinkage in equilibrium}
The analysis is started by considering the simplest possible case, i.e. a one-dimensional system with two grains occupying the intervals $(-\infty,0]$ and $[0,\infty)$.
The equilibrium phase-fields (vanishing $\frac{\delta F}{\delta \eta_\alpha}$) are given by
\begin{align}
\eta_1(x) &= 0.5 (1+\tanh(\frac{x}{W}))\\
\eta_2(x) &= 1-\eta_1(x)
\end{align}
with a uniquely determined interface width $W$ and the grain boundary at position $x=0$.
The grains are characterized by a constant density of $1$ and evaluating the chemical potential $\mu = \frac{\delta F}{\delta \rho}$ yields zero everywhere.
These results only consider the energy functional itself, without any influence from RBM.
If RBM is now introduced, there will be a net force and hence net velocity acting on the grains.
For the left grain $\eta_2$, the force density is described by $dF_2 = \kappa (\rho-\rho_{gb})g(\eta_2,\eta_1) (\nabla \eta_2 - \nabla \eta_1)$.
Since $\rho$ is 1 everywhere in equilibrium the density difference $\rho-\rho_{gb}$ is of positive sign.
Thus the direction of the force is initially given by $(\nabla \eta_2 - \nabla \eta_1)$ which points towards the bulk of $\eta_2$, i.e. in the negative x direction and away from the grain boundary.
By Newton's third law, the same but opposite force acts on the grain $\eta_1$ and hence both grains \emph{repel} each other.
This leads to the negative velocities observed in \cref{fig:velocity} which finally lead to the observed unshrinkage.
This conclusion can also be reached by considering that the force density will only vanish once $\rho = \rho_{gb}$ is achieved within the grain boundary, which necessitates transporting mass away from the grain boundary.
The implication of both arguments is also that generally the thermodynamic equilibrium state based on the functional is not the same as the equilibrium state in which the RBM term vanishes.
Hence nontrivial equilibrium states of this kind of model are generally dynamic with a spatially variable density and chemical potential field.
Only for the case $\rho_{gb} = 1$ the equilibrium state in 1D for both models overlap and hence a static equilibrium can be reached.

\begin{figure}[htb]
  \centering
  \includegraphics[width=0.95\columnwidth]{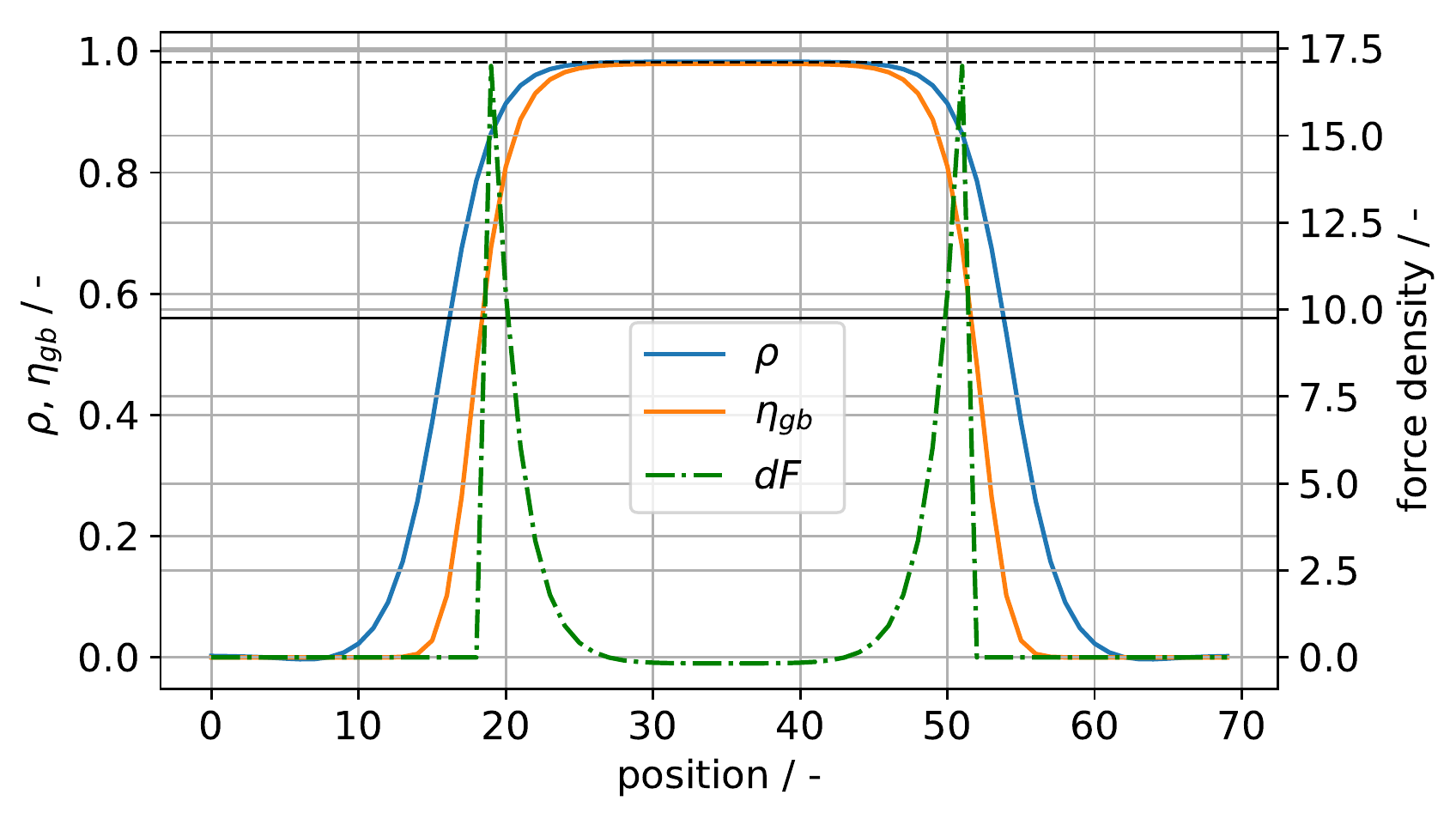}
  \caption{
  Density and grain boundary profile along the grain boundary for a two-particle simulation at $t=11.562$.
  The grain boundary $\eta_{gb}$ is defined as $4\eta_\alpha \eta_\beta$ in order to scale it in the range of $[0,1]$.
  The horizontal dashed black line indicates the chosen $\rho_{gb} = 0.9816$ and the horizontal solid black line the threshold value $c = 0.14$ multiplied by 4, accounting for the scaling.
  The force density $dF$ is non-zero only between the intersections of the solid black line and $\eta_{gb}$, as indicated by the dot-dashed line.
  Note that the region where $\eta_{gb} \sim 1$ exhibits only negative force densities.}
  \label{fig:densityprofile}
\end{figure}

This obviously raises the question of why there is enhanced densification with RBM in the simulations above, when the simplest case already shows unshrinkage.
The values of the density $\rho$ on the grain boundary are of key importance and hence the value range is explored in the following:
The two driving forces for a change in $\rho$ are the RBM flux, in equilibrium for a density of $\rho_{gb}$, as well as the diffusion flux, equilibriated at $\rho_{eq} = 1+f(\kappa)$ due to the Gibbs-Thomson effect slightly changing the equilibrium density.
Hence the density should lie within the interval $[\rho_{gb}, \rho_{eq}]$ as long as the considered point is within the grain boundary.
The grain boundary itself is attached to two triple points in the benchmark geometry and the density has to drop to 0 once the triple point is left behind and the pure vacuum is entered.
Thus regions within the triple point can have density values of $\rho < \rho_{gb}$.
The triple point regions need to be considered since the filtering function 
\begin{align}
   g(\alpha,\beta) &= \begin{cases}
                     1, & \eta_\alpha \eta_\beta \geq c\\
                     0, & else\\
                    \end{cases}
                    \label{eq:filterf}
\end{align}
can include them, and in fact does for the literature value of $c=0.14$.
Hence the interval is expanded to $[\rho_c, \rho_{eq}]$, with $\rho_c$ being the density at which $\eta_\alpha \eta_\beta = c$ holds.
An example of such a profile is plotted in \cref{fig:densityprofile} for the two-particle simulation with RBM at $t=11.562$, which still showed shrinkage.
Only the region between the intersection of the orange curve and the solid black line contributes to the total force, since values outside of it are filtered away by $g(\alpha,\beta)$.
This is visualized with the green line which corresponds to the local force density $dF$ except for the phase-field gradient.
The dashed black line indicates $\rho_{gb} = 0.9816$ and thus we can see that $\rho_c < \rho_{gb}$; this will generally hold unless one chooses $c$ very close to $0.25$ which is the maximum of $\eta_\alpha \eta_\beta$.
The parts of the profile above the dashed black line will cause repulsion of the particles, whereas those below it will cause attraction.
Thus if the middle region, the grain boundary, grows long enough the simulation will have net forces acting in a repelling manner.
This causes unshrinkage if the absolute value of the advection flux induced by these is larger than the diffusive flux, since the latter always acts in a densifying manner.
The volume of the particles enters the problem here, as the force is translated into a velocity with $\vec{v_{\eta_\alpha}} = \frac{m_t}{V_\alpha} \vec{F_\alpha}$ and thus larger particles are less likely to show unshrinkage.
In total determining the state when unshrinkage starts is not analytically tractable, as it depends on both time-and-space-dependent density and phase-field profiles as well as the global state via the particle volume.
\corM{If one only considers the velocity magnitude to be the determining factor for the influence of unshrinkage, one would conclude that larger particles would enter unshrinkage at later simulations stages.
This is due to the velocity magnitude being directly antiproportional to the particle size.
However, a larger particle size also implies reduced diffusional fluxes which always act densifying and hence would counteract unshrinkage.
As shown in the earlier study on particle size, the relative neck radius at the onset of unshrinkage in fact decreases as particle size increases.
This is likely due to the diffusional fluxes decreasing more than the advective fluxes when the particle size is increased.}

In order to exclude the possibility of unshrinkage, the force and thus velocity need to be kept attractive.
The direction of the force is affected by the choice of gradient vectors, the regions considered to be a grain boundary via $g(\alpha,\beta)$ and the difference $\rho-\rho_{gb}$.
Inverting the direction of the gradient vectors fails directly, as any initial contact between particles will force them apart and no shrinkage at all is possible.
Augmenting this by choosing the grain boundary region such that $\rho > \rho_{gb}$ is guaranteed might fix this approach, but it requires a precise calibration of the filtering parameter $c$ in \cref{eq:filterf} such that $\rho_c \geq \rho_{gb}$ is true.
A much easier calibration is available for the difference $\rho-\rho_{gb}$:
Since the the upper limit of $\rho$ within the grain boundary is known to large precision via $\rho_{eq} = 1+f(\kappa)$, the difference can be forced to be largely of negative sign which will lead to attractive forces.
This reduces the problem to finding the average curvature during the simulation or estimating it prior to the simulation if it is expected not to change appreciably.
More generally, as long as $\rho_{gb} \geq \rho_{eq}$ holds the grain boundary will tend to attract the adjacent grains.
Note that $\rho_{gb}$ is thus unrelated to the physical density of the grain boundary, but rather simply a parameter for ensuring attractive grain boundaries.

\begin{figure}
   \centering
  \includegraphics[width=\columnwidth]{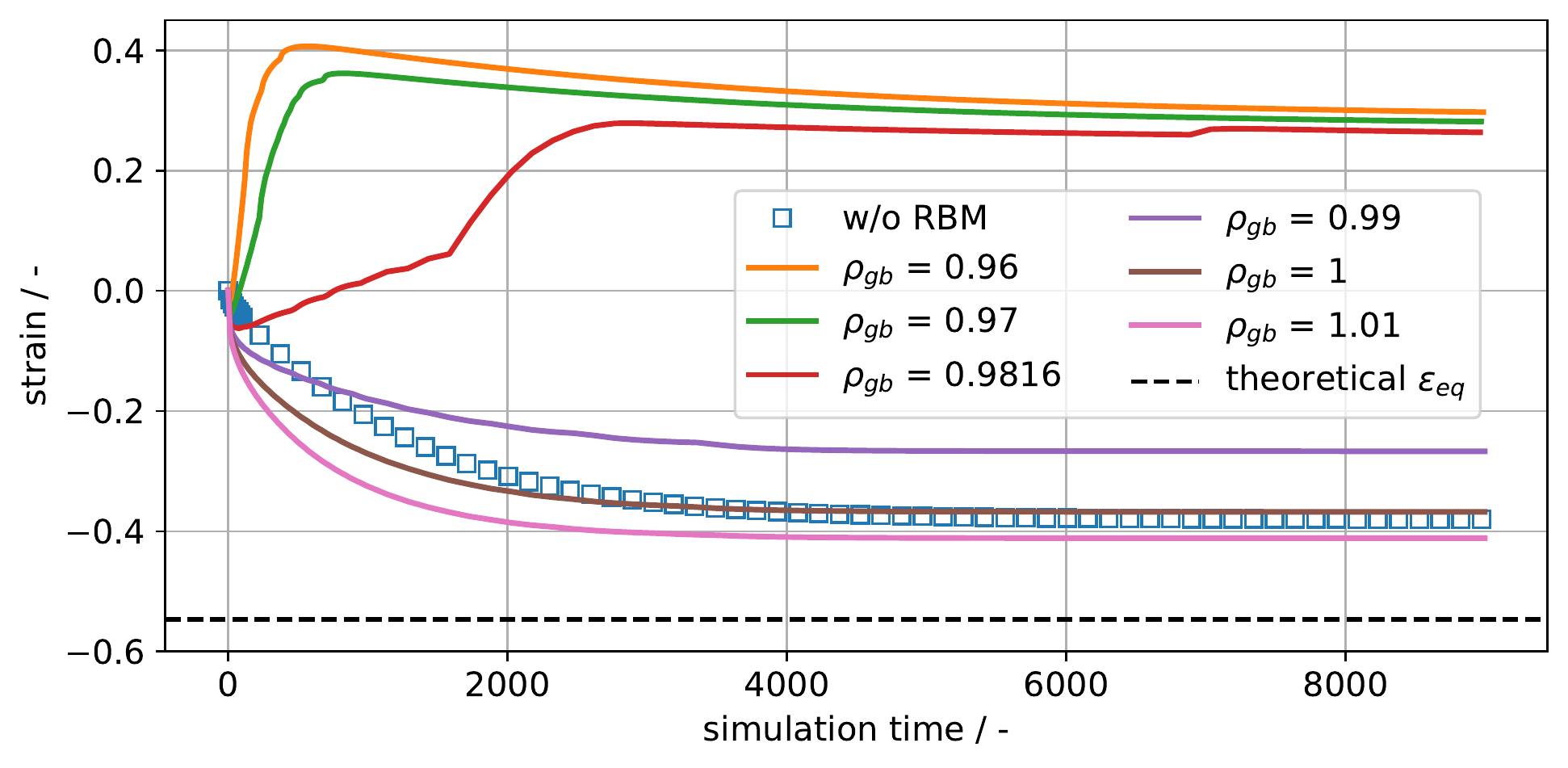}
  \caption{
  Strain of a two-particle system without RBM and with RBM for various grain boundary densities $\rho_{gb}$.
  \corM{A positive strain corresponds to a lengthening (unshrinkage) and a negative strain to a shortening (shrinkage).}
  For $\rho_{gb} \geq 0.99$ a monotonic behavior is observed as would be expected, but for values below $0.99$ unshrinkage is observed.
  The equilibrium strain is observed to depend on $\rho_{gb}$.
}
  \label{fig:c0var}
\end{figure}

In order to verify this procedure without needing to find $f(\kappa)$ for the given energy functional, several two particle-simulations with different values of $\rho_{gb}$ \corM{are carried out for up to 300 million time steps, corresponding to a simulation time of roughly 9000.
This value was sufficient to reach equilibrium for the simulations showing monotonic densification behavior.}
In general, the effect of the Gibbs-Thomson effect on the bulk density is small.
Hence grain boundary density values $\rho_{gb} \in \{0.96, 0.97, 0.9816, 0.99, 1.00, 1.01\}$ were chosen for this test.
Furthermore, one simulation without RBM was carried out to long times as well in order to compare the equilibrium grain boundary lengths and hence determine which model more closely approximates the densifying geometry from Kellet and Lange\cite{Kellett1989}.
The results of this simulation study are shown in \cref{fig:c0var}.
For values of $\rho_{gb} \geq 0.99$ monotonic behavior in the strain is observed, suggesting that for $\rho_{gb} \geq 0.99$ shrinkage is ensured.

\begin{table*}
\caption{\corM{Measured and theoretical equilibrium values for densifying simulations.}}
\hspace{-3cm} 
\label{tab:restab}
 \begin{tabular}{c | c c c c | c}
  property & no RBM & $\rho_{gb} = 0.99$ & $\rho_{gb} = 1.00$ & $\rho_{gb} = 1.01$ & theoretical\\
  \toprule
  shrinkage $|\epsilon|$ / - & 0.377 & 0.267 & 0.368 & 0.411 & 0.546 \\
  GB length $h$ / $r_i$ & 2.38 & 1.92 & 2.33 & 2.55 & 3.38 \\
  estimated $h'$ / $r_i$ ($\psi = \SI{150}{\degree}$) & 3.00 & 3.42 & 3.01 & 2.95 & 3.38 \\
  dihedral angle $\psi$ / $\si{\degree}$ & 159 & 145 & 157 & 160 & 150 \\
  estimated $h'$ / $r_i$ ($\psi$ measured)  & 3.42 & 3.23 & 3.38 & 3.44 & 3.38
 \end{tabular}

\end{table*}

\begin{figure}
   \centering
  \includegraphics[width=\columnwidth]{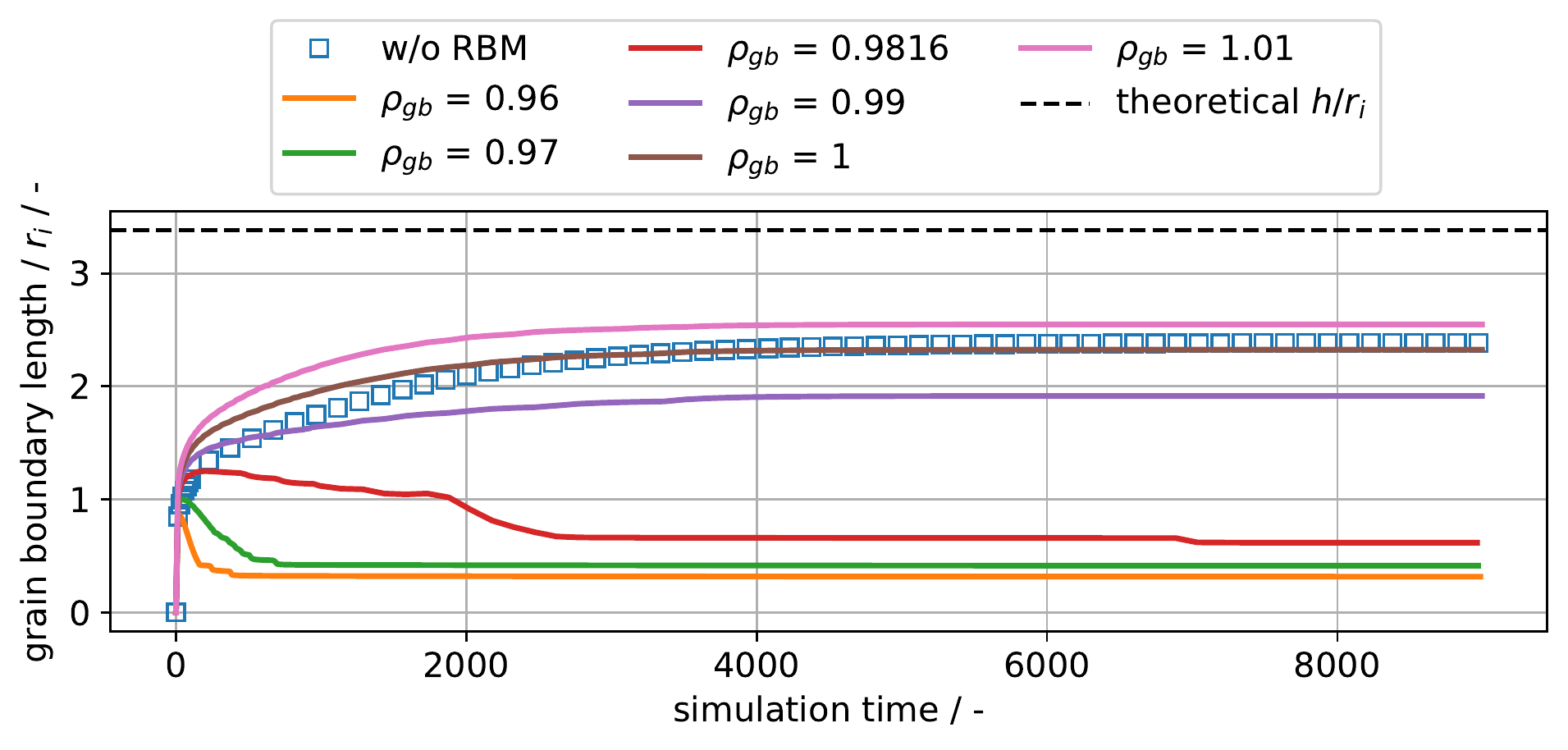}
  \caption{
  Grain boundary length $h$ normalized by the initial grain radius $r_i$ over time for various simulations.
  Like with the length change, increasing values of $\rho_{gb}$ exhibit longer grain boundaries, as both are correlated via mass conservation.
  Note that the simulation without RBM reaches an equilibrium length comparable to that of $\rho_{gb}=1$ and thus is still densifying.
}
  \label{fig:gblength_sim}
\end{figure}

\begin{figure}
    \centering
  \includegraphics[width=\columnwidth]{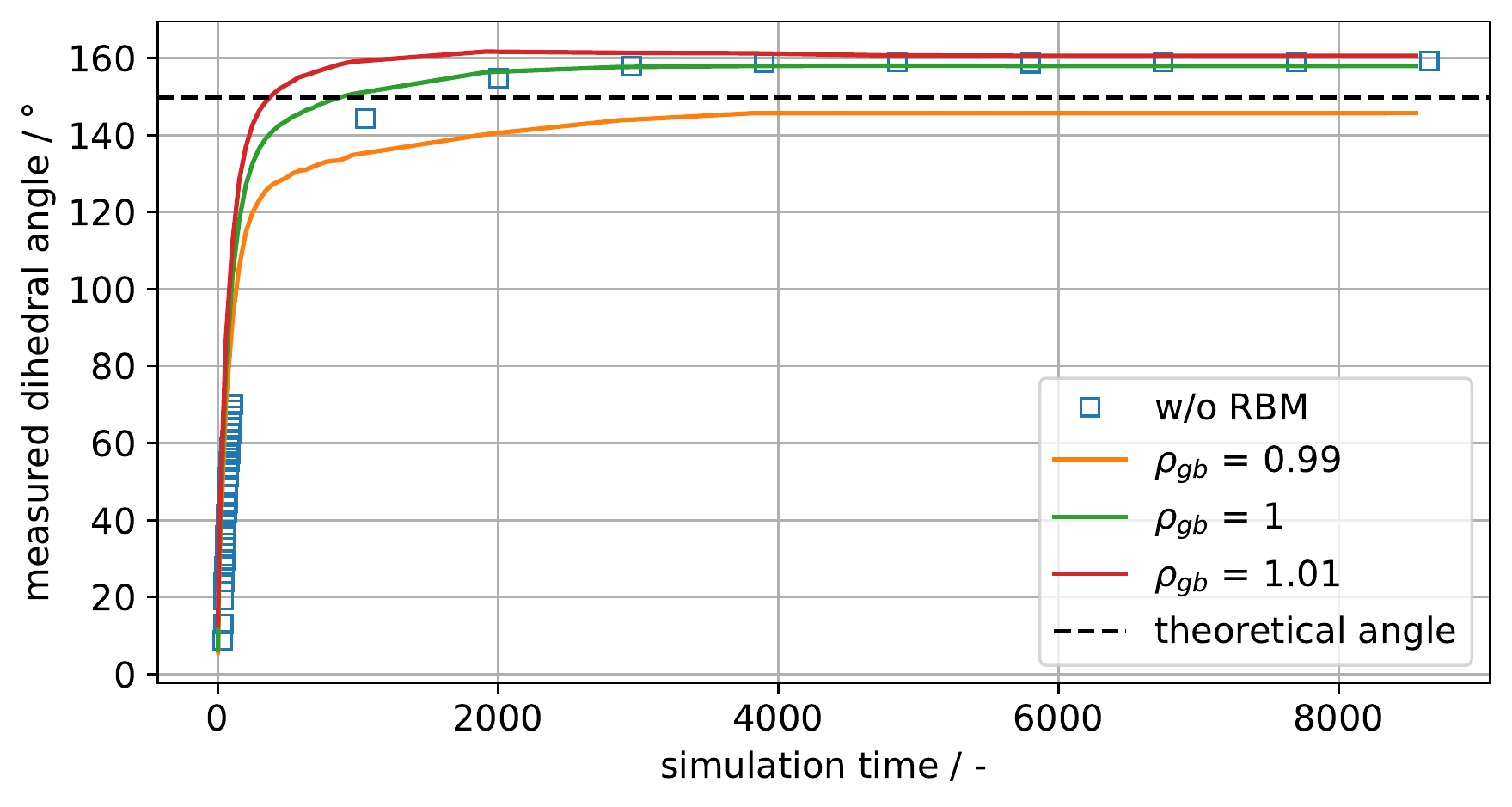}
  \caption{
  \corM{
  Dihedral angle of all densifying simulations as well as the theoretical value.
  There is a discrepancy for all simulations, with the grain boundary density influencing the dihedral angle.
  Note that this measurement is done on the field-resolved data and hence has less points than the previous plots which were calculated during the simulations.
  }
}
  \label{fig:angles}
\end{figure}

\corM{In the following, the influence of $\rho_{gb}$ on the shrinkage as well as its relation to the theoretical equilibrium shrinkage is discussed, with the results being collected in \cref{tab:restab} to give a concise overview.
As can be seen in \cref{fig:c0var}} different values of $\rho_{gb}$ lead to different equilibrium shrinkages, when it should be a universal value.
This difference is due to the proportionality of the force density to $\rho-\rho_{gb}$ and hence different values of $\rho_{gb}$ will directly change the equilibrium state.
Shrinkage values, corresponding to $-\epsilon$, of ${0.267, 0.368, 0.411}$ are observed for RBM simulations with $\rho_{gb} = {0.99, 1.00, 1.01}$ respectively, i.e. a higher $\rho_{gb}$ leads to larger shrinkage in equilibrium. 
The simulation without RBM achieved a shrinkage value of 0.377, comparable to that of $\rho_{gb} = 1.00$.
However, none of these values come close to expected infinite chain shrinkage at 0.546, very likely due to the finite length of the chain.
The same behavior and discrepancy is observed for the grain boundary length, which is plotted in \cref{fig:gblength_sim}.
Since shrinkage and grain boundary length are coupled via mass conservation, one may assume that these differences are correlated.
Hence an equivalent equilibrium grain boundary length $h'$, if the infinite chain could be simulated, can be estimated via $h' = h \frac{\epsilon_{infinite}}{\epsilon_{finite}}$.
For the simulations with $\rho_{gb} = \{0.99, 1.00, 1.01\}$ the grain boundary lengths $\{3.42, 3.01, 2.95\}r_i$ are calculated and for the simulation without RBM a length of $3.00r_i$ is calculated.
Thus the simulation which comes closest to the theoretical value of $3.38r_i$ is that for $\rho_{gb}=0.99$.
While it would be an interesting validation to derive an expression for the equilibrium grain boundary length and shrinkage for finite chains and compare them with the present results, it is out of the scope of this paper.
\corM{However, additional information can be gained from evaluating the dihedral angle in the simulations, as it enters the theoretical problem as a key value.
The results over time are shown in \cref{fig:angles}.
There is a deviation from the theoretical value even in equilibrium, but of similar magnitude as others' results\cite{Greenquist2020} for a dihedral angle of $\SI{150}{\degree}$.
It can be observed that as $\rho_{gb}$ is decreased, the dihedral angle is reduced.
Employing the earlier estimation of $h'$ with the observed dihedral angle yields the values $h' = \{3.23, 3.38, 3.44\}r_i$ for RBM simulations with $\rho_{gb} = \{0.99, 1.00, 1.01\}$ respectively and $h' = 3.42r_i$ for no RBM.
In this case the simulation with $\rho_{gb}=1$ matches the infinite chain result rather well.
Hence some of the earlier discrepancy is likely due to a different dihedral angle obtained in the simulation compared to theory.
The remaining difference in the shrinkage should thus mainly be due to the finite chain length.}

In any case, all of the simulations with monotonic shrinkage behavior, including the one without RBM, show grain boundary lengths somewhat comparable to the theoretical result for the densifying geometry.
Thus both models approximate the densifying geometry, with the RBM model only doing so if $\rho_{gb} \geq 0.99$ holds for the present setup.

\section{Discussion}
The previously shown simulation results and analytical considerations clearly show that some kind of advection term is necessary in order for phase-field models to exhibit sensible shrinkage behavior with more than two particles.
The physical background of this necessity is explored in the following:
Consider an atom or a vacancy located within an inner particle of the chain and its driving forces for migration.
The closer the atom is to a particular neck, the more likely it is to migrate towards it, with an atom perfectly between two necks not having any preferential direction.
This implies a symmetry of the mass flux towards both grain boundaries of an inner particle and thus no net movement of the particle center.
For the outer particles of the chain there is no such symmetry and thus they account for most of the motion observed in diffusion-only models.
In practice the motion of the outer particles causes a slight asymmetry for the inner particles, but as the simulation results have shown, this is negligible on diffusive timescales.
Adding an advection term whose velocity points on average to the total center of mass will induce a preferential direction for the mass flux towards the total center of mass of the system.
If the magnitude of this mass flux now depends on how many grain boundaries are crossed, then a velocity profile leading to a constant shrinkage independent of chain length can be established.
In effect the velocity field removes vacancies from the grain boundaries and neck regions and adds them on the opposing free surfaces, thereby annihilating vacancies within the grain boundaries and generating them again on the surface.
While this effect might be mimicked by a local source term, this would not lead to a preferential direction for the mass flux and thus is unlikely to show shrinkage independent of chain length.
Note that the advection term does not have to be based on a rigid-body motion model, it could also originate from solving a momentum transport equation.
Hence it would be interesting to verify whether phase-field models for liquid-phase and viscous sintering \cite{Villanueva2009,Yang2019} naturally include the proper scaling with chain length.
These models do not introduce additional parameters such as $\rho_{gb}$ or include external forces, but rather can be derived thermodynamically consistently as done by \cite{Yang2019} which ensures that the free energy is indeed minimized.

The second part of this paper elucidated the reasons behind the observed unshrinkage in simulations if they are continued long enough.
It was found that $\rho_{gb}$ is a parameter of key importance, as it controls whether grain boundaries act to repulse or attract the grains they are attached to.
Specifically, any value below $\rho_{eq}$ will force the grain boundaries to repulse the grains, but this may be balanced by the attractive force of the triple points.
This balancing is very likely the reason why the unshrinkage phenomenon has not been observed previously in phase-field simulations, as the state at which the balance tips towards repulsion occurs only late into the sintering process.
Based on the study on variation of $\rho_{gb}$, choosing $\rho_{gb} = \rho_{eq}$ is likely to yield the results closest to analytical theories of shrinkage.
In general $\rho_{eq}$ is a function of simulation state via the Gibbs-Thomson effect and hence should be estimated during the simulation run if the curvature is expected to change appreciably.

Finally, do all phase-field simulations of sintering, regardless of stage, require the inclusion of an advection term?
Certainly those that start from green bodies do, as the majority of densification still needs to occur without any kind of size dependence.
However, in the final stage only isolated pores remain and often these include gases which exert a pressure on the surrounding grain structure.
Assuming that these pores have reached an equilibrium pressure-size state, then the generation of vacancies on their surface would disturb the equilibrium and hence be energetically unfavorable.
Hence if the simulation is only concerned with pressurized, isolated pores, such as in \cite{hoetzer15-4,Joshi2017}, then including an advection term is unnecessary.

\section{Summary and conclusion}
In this work the necessity of including an advection term in phase-field models of sintering was shown by simulation.
Specifically a shrinkage rate independent of system size was only observed for models with an advection term.
Hence, in order to reproduce the correct kinetic scaling of sintering an advection term needs to be included.
Furthermore, a sensitivity study on the grain boundary density $\rho_{gb}$ showed that its choice is critical:
If $\rho_{gb}$ is chosen below the equilibrium grain density $\rho_{eq}$, then unshrinkage can occur.
The most practical choice of $\rho_{gb}$ is $\rho_{eq}$, as then the equilibrium states of the energy functional and the RBM model are very close and $\rho_{eq}$ can be calculated based on the energy functional.
In the study it could also be shown that regardless of whether RBM is included, simulations will approximate the equilibrium densifying geometries of analytical models\cite{Kellett1989}.

Future investigations into calculating velocity fields for solid-state sintering should verify the following points:
\begin{itemize}
 \item the thermodynamic equilibrium state is unmodified by the addition of the velocity field
 \item shrinkage is independent of system size
\end{itemize}
Given the present paper's investigation, the latter point is likely fulfilled by any model including RBM driven by grain boundary density difference.
However, these models are also plagued by the first point in that the state in which the RBM vanishes does not in general correspond to the thermodynamic equilibrium state.
\corM{It could also be shown that this difference in equilibrium states changes the dihedral angle, which now depends on $\rho_{gb}$ .}
Models for liquid-phase and viscous sintering such as \cite{Villanueva2009,Yang2019}, which are based on continuity equations, should not suffer from the first point, but whether they display constant shrinkage rates independent of system size needs to be investigated.
These models would need to be extended to incorporate the effect of vacancy generation and annihilation at grain boundaries.
A further point of interest would be including the effect of the grain boundary directly in the energy functional as done in \cite{Greenquist2020}.
This could independently solve the problem of unshrinkage since the thermodynamic equilibrium state is moved closer to that of the RBM model.


\section{Acknowledgment}
The author thanks Sumanth Nani  Enugala, Henrik Hierl and Michael Kellner from the Karlsruhe Institute of Technology and Fritz Thomsen from the Hochschule Flensburg for fruitful discussion.
\corM{The author also thanks the anonymous reviewer for pointing out the possibility of explaining the difference in strain obtained in simulations via the dihedral angle.}
The author gratefully acknowledges financial support by the Deutsche Forschungsgemeinschaft (DFG) under the grant number NE 822/31-1 (Gottfried-Wilhelm Leibniz prize), acquired by Prof. Britta Nestler of the Karlsruhe Institute of Technology.


\section*{Data availability}
The data and code employed in this paper are available upon reasonable request to the author.
\corM{The calculation of the equilibrium shrinkage following \cite{Kellett1989} is available at \url{https://git.scc.kit.edu/xt5201/rbm_supmat/-/tree/main/}.}

\bibliographystyle{unsrt}
\bibliography{literatur}


\end{document}